\documentstyle[11pt,paspconf]{article}

\begin{document}

\title{Mass loss and AGB evolution in Galactic satellites}

\author{Albert A. Zijlstra}
\affil{Department of Physics, UMIST, P.O. Box 88, Manchester M60 1QD, UK}



\begin{abstract}
The effect of metallicity on the AGB mass loss is reviewed.
Observations have mainly been limited to the Magellanic Clouds but are
observationally feasible throughout the Local Group.  Expansion
velocities are predicted to  depend strongly on $Z$ but the peak mass-loss
rates appear not to. The Mira $PL$ relation shows no evidence for a
$Z$ dependence, giving a powerful potential distance indicator. I
derive a distance modulus to the LMC of $m-M = 18.63 \pm 0.09$ as well
as a bright calibration for the Horizontal Branch in globular
clusters.  Finally, the predicted initial-final mass relation at low
$Z$ is shown to give rise to higher mass remnants. This would result
in an increased supernova rate in young, low-$Z$ populations, such as
found in the early Universe.
\end{abstract}




\section{Introduction}

AGB mass loss is generally believed to be a two-step process,
involving pulsation-driven atmospheric shock waves, followed by
radiation pressure on the dust driving the outflow.  The dust
production rate, and the eventual dust-to-gas ratio of the ejected
material, will depend on the metallicity of the star.  The details 
of the mass loss may therefore be different for different
stellar populations.

Existing theoretical formalisms for the mass are generally derived
from the Reimers mass loss: $ \dot M_R = -4\, 10^{-13} LR/M $. This
equation is adjusted to fit existing observations as well as
constraints such as the initial--final mass relations. Examples are
Bl\"ocker's law ($\dot M = -4.8\, 10^{-9} (LR/M) (L^{2.7}/M) $ ),
Vassiliadis \&\ Wood's (1993) formalism and the relation from Judge
\&\ Stencel ($ \dot M = (2.3+/-1.3)\, 10^{-14}\,
(R^2/M)^{1.43+/-.23}$). Because observations of AGB mass loss have
been largely limited to Galactic stars, these formalisms have limited
predictive power for different stellar populations. So far, only Bowen
\& Willson (1992) have attempted to predict mass-loss rates at low $Z$
from theoretical pulsation models.

The study of extra-galactic mass loss is thus of interest to extend
the parameter space of mass loss. At the same time, since the majority
of stars with life times less than a Hubble time will have passed
through the AGB, a better understanding of mass loss especially at low
$Z$ is also important for our knowledge on stellar remnants and on the
recycling of stellar gas into the ISM.   The lack of observational data is
mainly due to the difficulty to detect dust at any large distance, and
the paucity of 'dusty' AGB stars due to the short life time of the
superwind phase.

\section{Extra-galactic Observations}

\subsection{Capabilities}

Existing and planned instrumentation has the capability to detect AGB
stars to distances far outside the Local Group.  Observations of mass
loss of individual AGB stars is observationally possible at least for
the satellites of the Galaxy. Of these, studies of the Magellanic
Clouds have already been initiated (Reid 1991, Wood et al. 1992,
Zijlstra et al. 1996, van Loon et al. 1997,1998, Groenewegen \&\
Blommaert 1998).

Table 1 lists the observational limits of some present and future
instrumentation. The last column gives the maximum distance to which a
luminous AGB star may be detected given these limits. The K-magnitudes
refer to an unobscured, high-mass AGB star with K$=-9$. The N-band is
scaled from an obscured luminous AGB star in the LMC.  The NGST
value refers to low-resolution spectroscopy with $R=200$: the
value is very preliminary and only based on possible
instrumentation/telescope design.

\begin{table}
\caption{Obervational capabilities.} \label{tbl-1}
\begin{center}\scriptsize
\begin{tabular}{crrrrrrrrrrr}
\tableline
Band & Instrument/telescope & point-source sensitivity & distance limit \\
     &                      &  (5-$\sigma$ in 1 hour)  & (K$=-9$) \\
\tableline
K    &  SOFI/NTT            & 20.9 mag  &  10 Mpc \\
K    &  ISAAC/VLT           & 22.0 mag  &  16 Mpc \\
K    &  NGST                & 26.0 mag  &  100 Mpc $^a$ \\
N    &  TIMMI2/3.6m         & 2.5 mJy   &  1 Mpc $^b$ \\
N    &  VISIR/VLT           & 0.25 mJy  &  3 Mpc \\
Q    &  VISIR/VLT           & 10: mJy   &  0.5 Mpc \\
\tableline
\end{tabular}
\end{center}

\tablenotetext{a}{Low-resolution spectroscopy}
\tablenotetext{b}{Scaled from F(12mu)=1.0 Jy at the LMC}

\end{table}

\subsection{Individual stars}

The AGB population in the Magellanic Clouds is the only extra-galactic
population for which extensive observations are available. ISO
observations give the first reliable information on AGB mass loss and
are presented elswhere in these proceedings (van Loon, Blommaert et
al.).

AGB stars in more distant galaxies can be detected using infrared
arrays, however this is so far limited to non-mass-losing stars.
Davdige et al. (1998) present JHK images of the Sculptor group
galaxies, and detect indications for luminous AGB stars below K=16,
near the detection limit.  Alonso et al. (1998) present VIJHK data for
the dwarf galaxy NGC 3109. The AGB is clearly detected for ages down
to 1Gyr, but the survey is not deep enough to detect older AGB stars
in the infrared. The new generation of wide-field infrared arrays is
expected to give rapid progress in the next few years.

Evolved AGB stars can also be identified through their variability,
using I-band observations.  Tolstoy et al (1996) observed suspected
Cepheid variables in Gr-8 (d=2.4Mpc). One star in their sample (V-1)
both has a possible long period of 110 days and very red colours: It
is presently the best candidate for a Mira-like variable outside the
Local Group. 

Of the Galactic dwarf spheroidals, the presence of carbon stars is
indicative of intermediate-age populatios (younger than globular
clusters). Especially Fornax shows evidence for star formation less
than 2Gyr ago (Beauchamp et al. 1995) and may well contain mass-losing
stars.  Whitelock et al. (1996) study carbon stars in the nearby
Sagittarius dwarf: one star of their sample has since been shown to be a
carbon Mira. A review of extra-galactic carbon stars is given by
Groenewegen (these proceedings).

\subsection {Integrated properties}

Although individual mass-losing stars can only be detected at limited
distances, their integrated flux over a galaxy may be appreciable.
Knapp et al. (1992) studied elliptical galaxies in Virgo: they showed
that on average 40 per cent of their 12-micron emission may be due to
circumstellar dust. The expected photospheric contribution is
$S_{12}=0.08 S_{2.2}$ (with $S$ in Jy), whereas the observations for
Virgo ellipticals give $S_{12} = 0.138+/-0.014S_{12}$.  This is in
contrast to globular clusters where the 12-$\mu$m emission is
consistent with purely photospheric emission (Knapp et al. 1995, Penny
et al. 1997).

Recently, Bressan et al. (1998) have used the Padova isochrones to
predict integrated IR colours of stellar populations for various
mid-infrared bands.  Their results rely heavily on assumed
relations. e.g. between $Z$ and $\dot M$, but can easily be updated if
required.

\section{Expansion velocities at low metallicity}

The expansion velocity $v_{exp}$ of the gas is a result of the
radiation pressure on the dust, and the 'drag' or coupling between the
dust and the gas. This will depend on several parameters: $\dot M$ (or
the optical depth through the envelope), the stellar luminosity $L$
and the dust-to-gas ratio.  Habing et al. (1994) calculated outflow
velocities as function of these parameters and showed that $v_{exp}$
reaches a maximum for intermediate values of $\dot M$ but continuously
increases with $L$: $v_{exp} \propto L^{0.35}$. However, a very large
effect on the dust-to-gas ratio was found: the expansion velocities
drop to only a few km/s if the dust-to-gas ratio drops by a factor of
10 compared to AGB stars of solar metallicity, although the effect
differs in magnitude between carbon and oxygen-rich stars.

Few observations of expansion velocities outside the Galactic disk are
available.  Wood et al. (1992) determined $v_{exp}$ for five OH/IR
stars in the LMC. They found values of only 60\% of similar Galactic
OH/IR stars. This conclusion however depends on the way $v_{exp}$ is
measured: Wood et al. used the two strongest peaks in the OH spectrum
which may not be appropriate for noisy spectra. If instead the
outermost emission peaks are used, Zijlstra et al. (1996) find that
the four confirmed AGB stars have $v_{exp}=12$ km/s versus the Galactic
comparison stars $v_{exp}=14$ km/s, a difference of only 15\%. The
fifth star of Wood et al. is a luminous supergiant: SiO emission
showed that in this case the OH was not centred on the stellar
velocity and therefore $v_{exp}$ was greatly underestimated
(van Loon et al. 1996).

Groenewegen et al. (1997) detected CO for one of two mass-losing
carbon-Miras in the Galactic halo. The star has $v_{exp} = 3.2 \pm
0.2$ km/s. In comparison, Galactic disk stars average around 10 km/s
(Groenewegen et al. 1998). A metallicity of [Fe/H]$=-0.7$ is estimated
for the halo carbon star.  The very low $v_{exp}$ is qualitatively
consistent with the Habing et al. predictions, but further data on
halo stars would be required for any conclusions.

The present data suggests that $v_{exp}$ may be lower at subsolar $Z$,
but quantative information for a range of mass-loss rates and
luminosities is lacking. The difference between the LMC and the
Galaxy may be less than predicted by Habing et al. (1994), but this
could easily be due to compensatory effects, such as higher luminosity
and/or different pulsation properties in the LMC stars.

\section{Mass-loss rates at low metallicity}

The mass-loss formalisms mentioned in the Introduction do not include
a direct $Z$-dependence. As the temperature on the early AGB is
$Z$-dependent, this may indirectly introduce lower mass-loss rates at
lower $Z$.  (For the mass-losing part of the AGB we do not have data
on how the temperature scale depends on $Z$.)  On the other hand, the
formalisms were not designed for different metallicities and cannot be
used for predictions. If the mass loss is driven by radiation pressure
on dust, a possibly quite large effect could be expected.

There are a number of observations of mass loss at low $Z$. Up to 50
mass-losing AGB stars are now known in the LMC, with several having
sufficiently high mass-loss rates to have significant extinction even at
K. Zijlstra et al. (1996) conclude that for these stars, there is no
evidence that $\dot M$ in the LMC is systematically lower than in
the Galaxy. Also the SMC stars in their sample show no such effect.
Groenewegen \&\ Blommaert (1998) find evidence for further highly
obscured AGB in the SMC which also exhibit high mass-loss rates similar
to those of Galactic OH/IR stars.

Further data comes from Mira variables in globular clusters and in the
Galactic halo. Groenewegen et al. (1998) derive dust mass-loss rates
of a few $10^{-9} M_\odot yr^{-1}$ for two distant halo Miras with low
Z. Mira variables in 47 Tuc and Omega Cen have been detected at 12
$\mu$m and show similar mass-loss rates (Gillet et al. 1988, Origlia
et al., 1995, 1997). The corresponding gas mass-loss rates are of
order $10^{-6} M_\odot yr^{-1}$ (Origlia et al. derive much lower
values but these are based on a Reimers law. Their dust detection
implies a much higher rate.) Frogel \&\ Elias (1988) find from a more
complete sample an {\it average} $ \dot M > 10^{-7} M_\odot yr^{-1}$
for long-period variables in globular clusters with
[Fe/H]$>0.1$. Thus, variables in globular clusters show mass-loss
rates within normal ranges. It should be noted that a correction may
have to be made if the expansion velocities are very low.

The data, within the limited accuracy, shows no evidence for low
mass-loss rates at low $Z$. This may be related to the fact that an
AGB star will continue to evolve to higher luminosities untill the
mass-loss rate exceeds the nuclear-burning rate. If mass loss is less
efficient at low $Z$, the star will reach higher luminosity which will
again increase the mass-loss rate (Habing et al. 1994). Thus, stars in
different populations are likely to reach similar mass-loss rates, but
may do so at different phases in their evolution (Bowen \&\ Willson
1991).

\section{Mira Period--Luminosity relations}

An important result obtained from extra-galactic AGB stars is the
existence of a well-defined period--luminosity relation for Miras in
the LMC. The relation was discovered by Glass \&\ Evans (1981). Feast
et al. (1989) showed that the relation is especially narrow (1-$\sigma
= 0.13$mag) when using the K-band.  Not all Miras follow the $PL$
relation. LMC Miras with periods longer than 400 days tend to
be overluminous (e.g.  Zijlstra et al. 1996). In contrast, OH/IR stars
near the Galactic Centre appear underluminous (Blommaert et al. 1998,
Wood et al. 1998), although there are uncertainties in the bolometric
corrections. A few nearby Miras are also underluminous (van Leeuwen et
al. 1997).

Bedding \&\ Zijlstra (1998) show that nearby Miras and semi-regulars
(SR) with accurate Hipparcos parallaxes define three sequences: One
sequence agrees with the LMC Mira $PL$ relation, a broader sequence
contains SR variables which is shifted by about a factor of two
towards shorter periods (e.g. Wood \& Sebo 1996), and a few stars fall well
below the relation.  (A number of semiregulars show two periods, one
of which agrees with the Mira relation and the other with the less
narrow SR sequence.)  The few stars below the relation have
luminosities similar to the semiregulars (although having much longer
periods) and may be evolutionary related to this group. The Miras show
slightly higher luminosities and appear therefore more evolved.  The
luminosity difference between SR's and Miras is consistent with an
evolution along a Whitelock (1986) track. From the Hipparcos data,
life times of the order of $5\, 10^5$ yr can be estimated for the Mira
phase and similar or larger for the SR's.

Wood (1990) predicts that the $PL$ relation should depend on $Z$, with
galactic Miras being less luminous at K by 0.25$\,$mag compared to LMC
Miras of the same period (or, Galactic Miras have longer periods).
This is however not confirmed by available observations (e.g., van
Leeuwen et al. 1997, Feast 1996, Whitelock et al. 1994).  Alvarez et
al. (1998) find some evidence for an offset between the Hipparcos
Miras and the LMC $PL$ relation, but the more accurate data of Bedding
\&\ Zijlstra (1998) does not confirm this.

\begin{figure}
\vspace{18 cm}
\caption{The Mira PL relations. Plus signs indicate LMC Miras (Feast
et al. 1989), circles Galactic Hipparcos Miras (Bedding \&\ Zijlstra
1998) and triangles Miras in globular clusters (Whitelock 1986).
LMC Miras are scaled to a distance modulus of 18.57. 
Three different distance scales for globular clusters are shown, based
on different calibrations for the horizontal branch magnitude.
 }
\label{fig-1}
\end{figure}

Figure 1 compares the LMC Miras with Hipparcos Miras and with Globular
Cluster Miras. The Hipparcos Miras (open circles) all have parallax
uncertainties less than 15\%, and the resulting Lutz--Kelker bias is
less than 0.05 mag (Koen 1992). A direct fit between the LMC and
Hipparcos Miras (applying a bias correction according to Koen 1992)
gives an LMC distance modulus of $18.63 \pm 0.09$, not inconsistent
with other determinations.  A shift of 0.25 mag would put the LMC at
an unlikely distance of $18.9$ and seems excluded.

Globular Clusters are shown in Figure 1 as triangles. All occur in
clusters with $[Fe/H]>0.1$ (Whitelock 1986).  To scale the data, two
Horizontal Branch distance scales of Harris (1996) are used, a short
scale (derived from RR Lyare statistical parallaxes) and a long one
(field subdwarfs).  The average scale corresponds to $M_V(HB) = 0.15
[Fe/H]+0.85$.  The bottom panel uses the Chaboyer (1998) relation
$M_V(HB) = 0.23 ([Fe/H]+1.6)+0.46$ which again is 0.1mag brighter than
an average including the statistical parallax. It is clear that only
the brightest scale gives good agreement with the Mira $PL$ relation.  A
direct fit to the LMC data using the Harris scale gives $M_V(HB) = 0.15
[Fe/H]+0.60\pm0.03$. (This would yield globular cluster ages of 11
+/-2 Gyr.)

There is thus little evidence for a dependence of the $PL$ relation
on $Z$, when using K-band magnitudes. If the relation is indeed
invariant, it would make Mira variables a highly useful
distance indicator.

\subsection{Mass loss and the PL relation}

The postion of a Mira in the $PL$ diagram is determined by the
core-mass--luminosity relation and by the pulsation equation:

\begin{equation}
\log P = 1.5 \log  R -0.5 \log M + \log Q
\end{equation}

\noindent where $P$ is the period in days, $R$ and $M$ are the radius
and mass in solar units and $Q$ is the pulsation constant. Using the
AGB relation of Wood (1990):

\begin{equation}
 M_{\rm bol} = 15.7 \log T_{\rm eff} +1.884 \log z -2.65 \log M
-59.1 -15.7\Delta 
\end{equation}

\noindent (where the disturbance term $\Delta$ is usually ignored), one finds
(Feast 1996):

\begin{equation}
   M_{\rm bol} = -2.036 \log P +0.73 \log z 
  - 2.049 \log M +2.881 +2.036 \log Q 
\end{equation}

Here $z = Z/Z_\odot$. The existence of a narrow $PL$ relation implies
$P$ is not an independent parameter: $ M_{\rm bol} = f(z,M)$.  Note
that $z$ is not {\it a priori} known for an LMC star but may vary
between stars of different ages. $M$ is not constant, due to mass
loss. In addition, $M_{\rm bol}$ is {\it independently from $M$ }
determined by the core mass, but will change by as much as 1 mag over
the phase of the thermal-pulse cycle. These separate effects should
lead to a large broadening of the $PL$ relation (e.g. Vassiliadis \&\
Wood 1993).  From the narrow $PL$ relation we find:

\begin{equation}
 dP/dM = 0;~~~~ dP/d M_{\rm bol} = -3.0; ~~~~ dP/dz = 0,
\end{equation}

\noindent valid during the mass loss phase. (The second relation is
the observed $PL$ relation.) To keep a star on the $PL$ relation
during the Mira phase, the stellar radius must adjust to compensate
for the changing luminosity and mass. The AGB relation above, valid
for the earlier AGB, may thus not fully describe Miras.

The observed radius is wavelength dependent, due to molecular bands in
the extended atmosphere. This re-introduces a metallicity effect, seen
as a colour term dependent on $z$ (Feast 1996). The $PL$ relation is
best defined at K, where the atmospheric bands do not contribute and
the observed radius is closest to the photospheric radius.

A possible explanation for the relations above is a feedback
between the stellar pulsations and mass loss. Such a mechanism could
break down either at high luminosities or at high mass-loss rates, where
the latter could correspond to the underluminous stars at the Galactic
centre.

\section{Initial--Final Mass relations at low Z}

\begin{figure}
\vspace{10 cm}
\caption{Left: initial-final mass relations for intermediate-mass
stars at low $z$. Right: The inferred relative number of type-II
(post-main-sequence) supernovae as function of $z$, integrated
over the first Gyr.} \label{fig-2}
\end{figure}

The AGB mass-loss formalisms include an $R$-dependence. Using the
$R\propto Z^{0.088}$ relation of Iben (1984) for the AGB predicts that
a low-$z$ AGB star will have a slightly lower $\dot M$ at the same
luminosity.  For $z=0.1$, the Reimers mass loss would be 20\%\ lower.
To compensate, a low-$z$ star would require a higher luminosity (and
therefore a higher core mass) to reach the same $\dot M$. Assuming the
evolution is terminated when $\dot M$ exceeds the nuclear burning
rate, a low-$z$ star is expected to reach a higher final core mass and
thus yield a more massive white dwarf than a star with the same
initial mass but higher $z$.

For $z < 0.1$, the effect of radiation pressure of dust becomes very
small and the mass loss is probably purely driven by the stellar
pulsation. For such a dustless wind, Bowen \&\ Willson (1991) find
that $M_{\rm bol} $ should be 0.3 mag brighter to reach the same mass
loss as a dusty wind. The total increase in brightness as function of
$Z$ to reach the same mass-loss rates is $- \Delta M_{\rm bol} = 0.30
-0.33 \log z$, for $z = Z/Z_\odot < 0.1$.

The core mass--luminosity relation $L = 59250(M_c-0.495)$ (e.g.
Boothroyd \&\ Sackman 1988) can be used to quantify the effect on the
final mass of the resulting white dwarf. The required increase is very
small for low core mass, but becomes significant for core masses
corresponding to the heaviest white dwarfs.  We use an
approximate present initial--final mass relation of $M_f =
0.5+M_i/12$, and a Salpeter IMF $dN/dM = m^{-2.35}$. Figure 2 shows
the resulting $M_i$--$M_f$ relation as function of $z$.  At $z<0.01$,
the heaviest remnants reach the Chandresekhar mass giving rise to a class of
AGB supernovae. At the extreme  $z$, the contribution of these AGB-SN
exceed the type II SN by a factor of a few, over the first Gyr of the
stellar population.  If confirmed, such an effect would have dramatic
consequences for the evolution of the earliest stellar populations
in the high-redshift Universe.

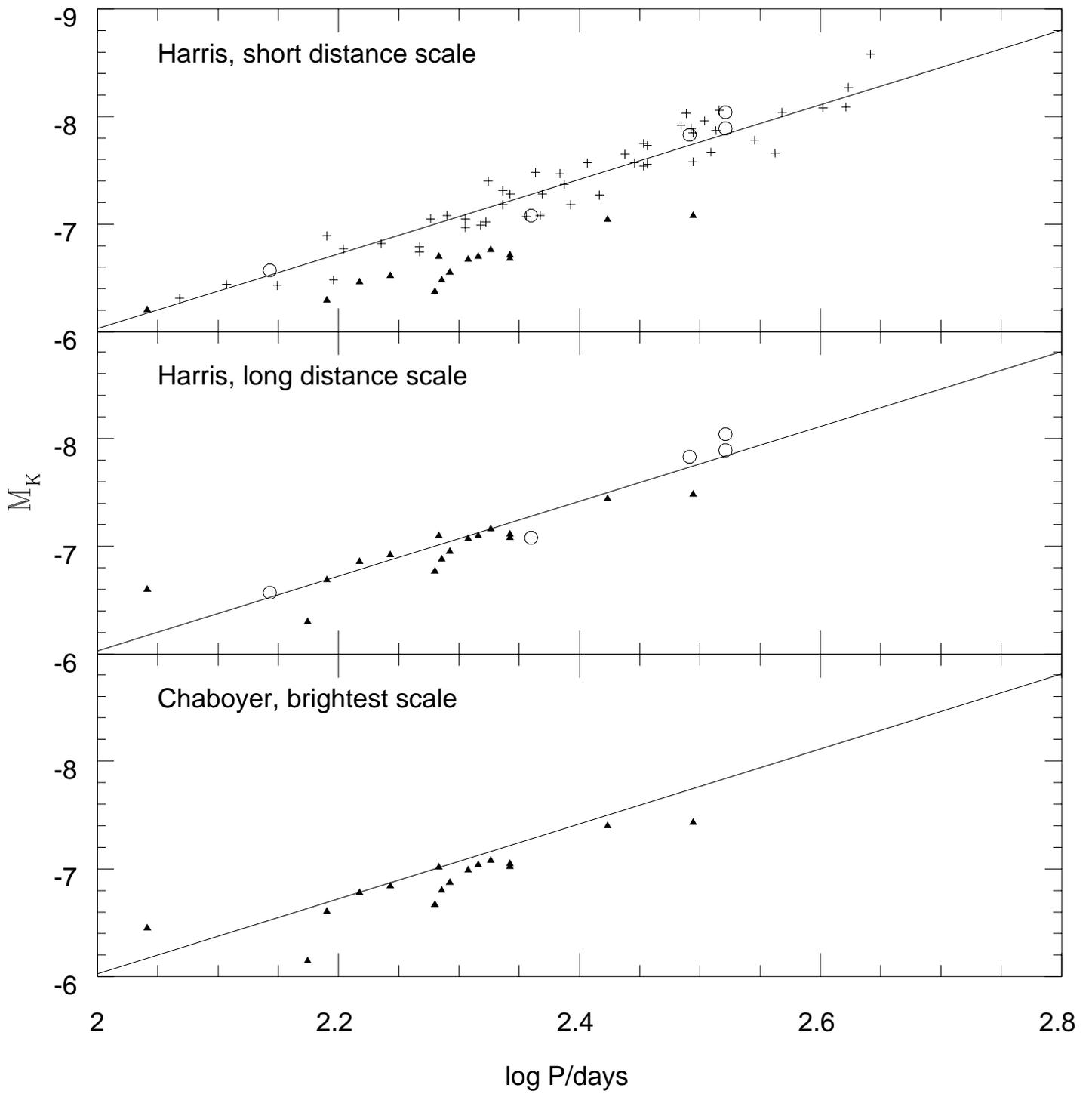
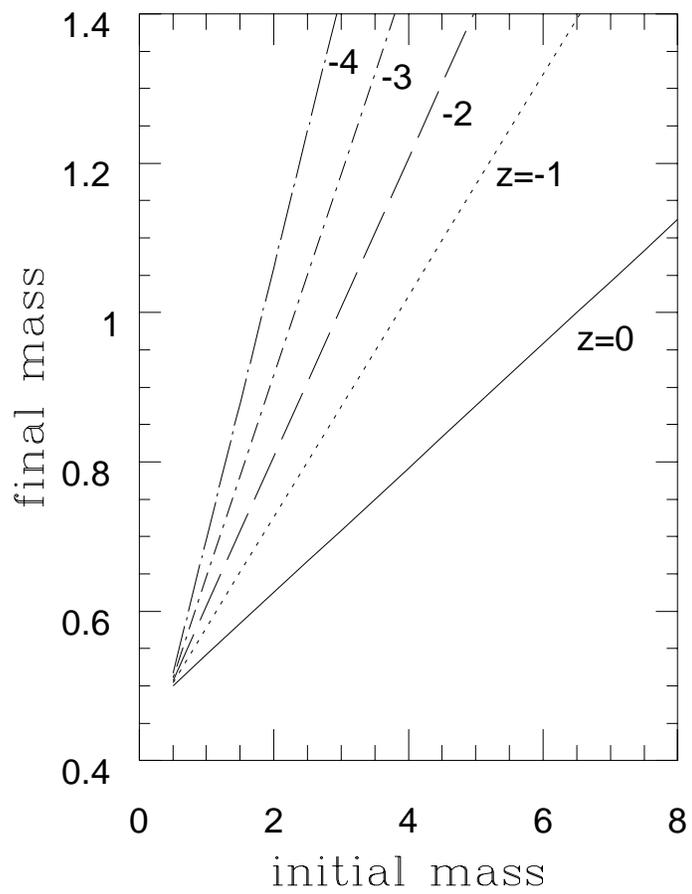
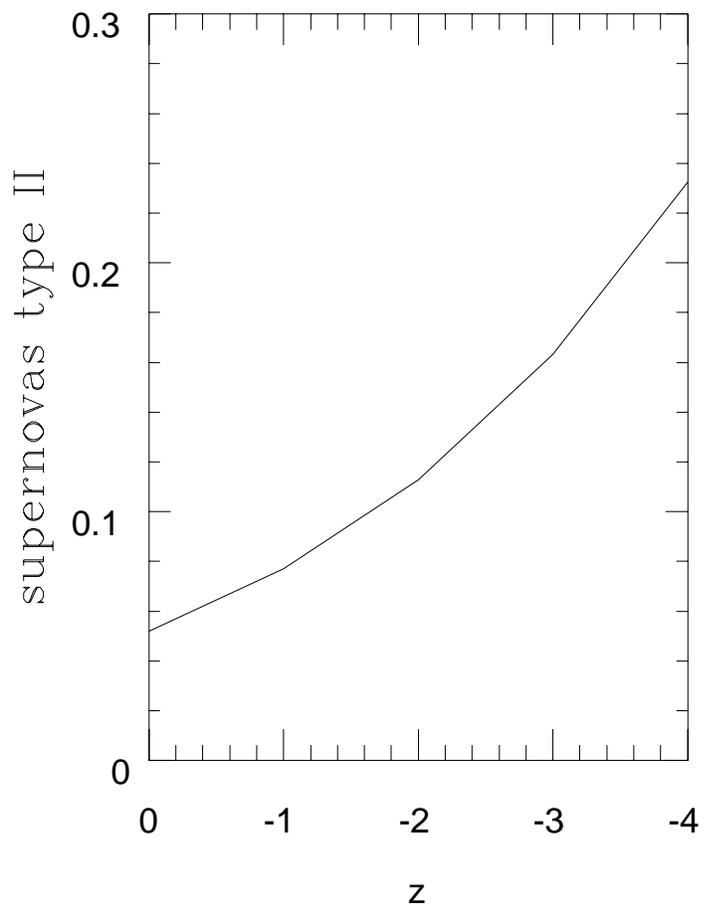
\end{document}